**Enderle *et al.* Reply:** In their Comment [1], Drechsler *et al.* argue that the quasi-1D spin-1/2 material LiCuVO$_4$ is characterized by a ferromagnetic nearest-neighbor exchange $J_1 < 0$ that is *stronger* than the antiferromagnetic next-nearest neighbor exchange $J_2 > 0$ (in contrast to our common earlier work where $|J_1|/J_2 < 1$ [2]), and that therefore neither the RPA model of our Letter [3] nor the observation of a 4-spinon continuum are valid.

Our RPA approach (a 1D model) is valid in the weak-coupling regime where $J_1$ is "small". Since the frustrated ground state of the $J_1$–$J_2$ chain is stable in the range $0 < -J_1/J_2 < 4$, "small" means consequently $|J_1|/J_2 < 4$. On a more quantitative level, evidence for the validity of the RPA is provided by the calculations presented in the Comment [1]. Figure 1(a) displays the energy $\Omega/J_2$ of the bound state at $k=1/4$ as a function of $|J_1|/J_2$. The RPA has the same behavior as DMRG and ED calculations in the whole range of $J_1/J_2$ of interest. Furthermore, the ED calculations of the dynamic susceptibility (Figs. 1(b) and 1(c) of [1]) display a two-spinon (2SP) like continuum characteristic of a $J_1=0$ Heisenberg antiferromagnetic chain, the starting point of the RPA. The ED calculations thus justify that this picture remains valid up to values of $|J_1|/J_2$ much larger than used in the RPA of [3].

The ratio $|J_1|/J_2$ cannot be determined from the bound state energy $\Omega$ alone, as correctly acknowledged in the Comment. However, using the observed neutron scattering intensities over the whole wave-vector range allow the determination of $J_1$ and $J_2$ independently, as was indeed done in [3]. Fitting $J_1$ for various fixed $J_2$ shows that there is a clear optimal value of $|J_1|/J_2$ [Fig. 1(b)].

It is considerably more difficult to extract exchange integrals with precision from macroscopic measurements, in particular when the exchange is frustrated as in LiCuVO$_4$. The analysis in [1] illustrates that a large difference in $|J_1|/J_2$ leads to changes in $M(H)$ at the percentage level, i.e. similar to experimental variations due to hysteresis, field direction, and pulse heating. Their 1D $T=0$ analysis also neglects the interchain interactions and the effect of temperature, both of which influence $M(H)$ of LiCuVO$_4$ [4], and apparently uses a too small $H_{\text{sat}}$ value, so that the experimental $M(H)$ is only 80% of the saturation value at their $H_{\text{sat}}$. Their resulting parameter set fails to reproduce the negative Curie-Weiss constant of magnetic susceptibility measurements [2, 5], in contrast to the 3D exchange parameters of Ref. [2] (line 3 in Table 1) which also lead to a reasonable $\mu_0 H_{\text{sat}}^c \approx 44$ T. The 1D RPA parameters of [3] agree well with this 3D set.

Independent of the choice of exchange parameters, the upper boundary of a two-particle continuum is given by twice the largest onset energy. In LiCuVO$_4$, the largest observed onset energy (at $k=0.74$) is less than 6 meV, which limits 2SP continuum states to below $12|\sin \pi k|$ meV. However, Fig. 4 of [3] demonstrates considerable spectral weight *above* $12|\sin \pi k|$ meV for $k > 0.5$, which can only come from more-than-2-spinon states, and we therefore attribute this intensity to 4-spinon excitations.

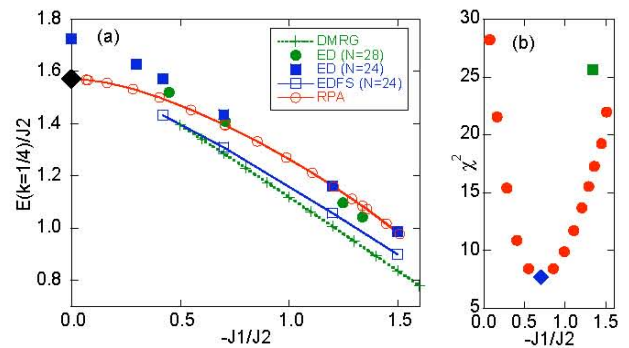

FIG. 1: (a) Bound-state energy at $k=1/4$ for various approaches: DMRG and Exact Diagonalization (ED) for 28 sites from [1], ED for 24 sites with and without finite-size extrapolation (this work), and RPA [3]. The solid diamond shows the exact $J_1=0$ limit. (b) Sensitivity of RPA fits with respect to $-J_1/J_2$ expressed as $\chi^2$ values. The green square and the blue diamond are the parameter sets of [1] and [3], respectively.

Certainly a unique quantitative determination of the exchange couplings in LiCuVO$_4$ is needed for a complete description. However, the absence of precise values for all parameters does not influence the main qualitative conclusions in [3], namely the experimental observation of a two-spinon continuum in a frustrated magnetic chain and clear indications of a possible four-spinon continuum.

M. Enderle[1], B. Fåk[2], H.-J. Mikeska[3], and R. K. Kremer[4]

[1] Institut Laue Langevin, BP156, 38042 Grenoble, France
[2] INAC, SPSMS, CEA, 38054 Grenoble, France
[3] ITP, Leibniz Universität, 30167 Hannover, Germany
[4] Max-Planck Institute, 70569 Stuttgart, Germany

[1] S.-L. Drechsler *et al.*, arXiv:1006.5070v2.
[2] M. Enderle *et al.*, Europhys. Lett. **70**, 237 (2005).
[3] M. Enderle *et al.*, Phys. Rev. Lett. **104**, 237207 (2010).
[4] L. E. Svistov *et al.*, (JETP Lett.) Pis'ma v ZhETF **93**, 24 (2011); arXiv:1005.5668.
[5] Hyun-Joo Koo *et al.*, arXiv:1103.1616.